\documentclass{PoS}

\title{First order transition regions in the quark masses and chemical potential parameter space of QCD}

\ShortTitle{First order transition regions in the quark masses and chemical potential parameter space of QCD}

\author{\speaker{Shinji Ejiri}\\
        Department of Physics, Niigata University, Niigata 950-2181, Japan\\
        E-mail: \email{ejiri@muse.sc.niigata-u.ac.jp}}
\author{Norikazu Yamada\\
        KEK Theory Center, Institute of Particle and Nuclear Studies, 
        High Energy Accelerator Research Organization (KEK),
        Tsukuba 305-0801, Japan\\
        School of High Energy Accelerator Science, The Graduate University
        for Advanced Studies (Sokendai), Tsukuba 305-0801, Japan
        }
\author{Hiroshi Yoneyama\\
        Department of Physics, Saga University, Saga 840-8502, Japan
}

\abstract{
We investigate the phase transitions of $(2+N_{\rm f})$-flavor QCD, where two light flavors and $N_{\rm f}$ massive flavors exist, aiming to understand the phase structure of (2+1)-flavor QCD. 
Performing simulations of 2-flavor QCD with improved staggered and Wilson fermions and using the reweighting method, we calculate probability distribution functions in the many-flavor QCD. 
Through the shape of distribution functions, we determine the critical surface terminating first order phase transitions in the parameter space of the light quark mass, heavy quark mass and the chemical potential, and find that the first order region becomes larger with $N_{\rm f}$.
We then study the critical surface at finite density for large $N_{\rm f}$ and the first order region is found to become wider with the increasing chemical potential. 
On the other hand, the light quark mass dependence of the critical mass of heavy quarks seems weak in the region we investigated. 
The result of this weak dependence suggests that the critical mass of heavy quark remains finite in the chiral limit of 2-flavors and there exists a second order transition region on the line of the 2-flavor massless limit above the tri-critical point.
Moreover, we extend the study of 2-flavor QCD at finite density to the case of a complex chemical potential and investigate the singularities where the partition function vanishes, so-called Lee-Yang zeros. 
The plaquette effective potential is computed in the complex plane. 
We find that the shape of the effective potential changes from single-well on the real axis to double-well at large imaginary chemical potential and the double-well potential causes the singularities.
}

\FullConference{9th International Workshop on Critical Point and Onset of Deconfinement\\
		 17-21 November, 2014\\
		 ZiF (Center of Interdisciplinary Research), University of Bielefeld, Germany}

\begin{document}

\section{Introduction}
\label{sec:intro}

QCD at high temperature and density has rich phase structure, and 
the nature of the phase transition changes depending on the quark mass and the number of flavors.
The quark mass dependence of the QCD phase transition is important not only for the theoretical interest in the phase structure of QCD with various quark masses but also for the investigation of finite density QCD or QCD near the chiral limit, in which the numerical studies are difficult.
From the numerical study of (2+1)-flavor QCD, the QCD phase transition is considered to be crossover at low density but is expected to turn into a first order transition at high density. Finding the endpoint of first order transition is then crucial for establishing the above expectation. 
Unfortunately, it is still extremely difficult to simulate quarks with physical mass at high density. But accessing the endpoint becomes easy if one extends QCD in appropriate directions, and importantly if such an extension is smooth, one can extract information on original QCD by suitable extrapolations \cite{crtpt,dFP03}.

Moreover, the nature of the phase transition in the chiral limit of 2-flavor QCD is a long-standing problem. 
The standard expectation is of second order, but the order is not conclusive due to the difficulty of the numerical study in the chiral limit.
One of the good approaches is to investigate the boundary of the first order transition region. 
If the critical value of the strange quark mass does not go to infinity in the up and down quarks massless limit, the transition of the massless 2-flavor QCD is not of first order.
However, it is difficult to study it, because the first order region is very small in (2+1)-flavor QCD and simulations with very small mass are required.

In Ref.~\cite{yamada13}, the boundary of the first order region is studied in many-flavor QCD, motivated by a feasibility study of the electroweak baryogenesis in technicolor theories constructed by many-flavor gauge theory.
They studied QCD with two degenerate light quarks of the mass $m_{\rm l}$ and the chemical potential $\mu_{\rm l}$ and $N_{\rm f}$ massive quarks with $m_{\rm h}$ and  $\mu_{\rm h}$, 
and found by measuring probability distribution function that the critical massive quark mass becomes larger as $N_{\rm f}$ increases. 
Therefore, the investigation of the critical line becomes easier as $N_{\rm f}$ increases. 
This extension of QCD is also useful for the study of the phase transition of massless
2-flavor QCD. If the critical mass for the massive $N_{\rm f}$ flavors remains finite in the massless limit of two light flavors, it gives a strong support for the second order transition.
A similar approach has been tried in Ref.~\cite{bonati14}.

This paper consists of two parts.
The first part deals with the light quark mass dependence of the critical massive quark mass separating the first order and crossover regions in $(2+N_{\rm f})$-flavor QCD. 
We perform simulations with 2 flavors of improved Wilson quarks. 
The effect from the dynamical $N_{\rm f}$-flavors are added by a reweighting method assuming that the $N_{\rm f}$-flavors are heavy. 
We then investigate the critical heavy quark mass as a function of the light quark mass through the shape of distribution functions and discuss the nature of phase transitions in the chiral limit of 2-flavor QCD.

The second part is the chemical potential dependence. 
In particular, we extend the real chemical potential $(\mu)$ to complex value.
The phase transition of 2-flavor QCD with finite mass at $\mu=0$ is crossover.
But, when the chemical potential is introduced, the shape of the distribution changes and the distribution function will become first-order-transition-like.
For the complex $\mu$, this property make a singularity, which is called ``Lee-Yang zero'' \cite{LeeYang}. 
Lee and Yang proposed the method to investigate the nature of phase transitions from the singularities in the complex parameter plane, and applications to finite density QCD have been discussed in Refs.~\cite{steph06,yoneyama13}.

In the next section, we explain our method to identify the nature of phase transitions via the distribution function. We then argue the light quark mass dependence of the endpoint of the first order transition in Sec.~\ref{sec:heavy}. 
The singularities in the complex plane are discussed in Sec.~\ref{sec:singularity}.
Conclusions are given in Sec.~\ref{sec:summary}.

\section{Critical point by a histogram method}
\label{sec:method}

We study QCD with two degenerate light quarks of the mass
$m_{\rm l}$, the chemical potential $\mu_{\rm l}$ and $N_{\rm f}$ heavy quarks.
We define the probability distribution function of average plaquette value, 
\begin{eqnarray}
w(P; \beta, m_{\rm l}, \mu_{\rm l}, m_f, \mu_f) 
&=&  \int {\cal D} U {\cal D} \psi {\cal D} \bar{\psi} \
\delta(P- \hat{P}) \ e^{- S_q - S_g} \nonumber \\
&=& \int {\cal D} U \ \delta(P- \hat{P}) \ 
e^{6\beta N_{\rm site} \hat{P}}\ (\det M(m_{\rm l}, \mu_{\rm l}))^2 
\prod_{f=1}^{N_{\rm f}} \det M(m_f, \mu_f)
\label{eq:pdist}
\end{eqnarray}
where $S_g$ and $S_q$ are the actions of gauge and quark fields, respectively, and 
$M$ is the quark matrix. 
For simplification, we adopt $M$ which does not depend on $\beta$ explicitly. 
$N_{\rm site} \equiv N_{\rm s}^3 \times N_t$ 
is the number of sites and $\beta=6/g_0^2$ is the simulation parameter.
$\hat P$ defined by $\hat P=-S_g/(6N_{\rm site} \beta)$ 
is $1 \times 1$ Wilson loop for the standard plaquette gauge action.
$\delta (P - \hat{P})$ is the delta function, which constrains the operator $\hat{P}$ to be the value of $P$. 
We moreover define the effective potential,
$V_{\rm eff}(P;\beta,m_f,\mu_f) = -\ln w(P;\beta,m_f,\mu_f)$.

Denoting the potential of 2-flavor at $\mu =0$ by $V_0(P; \beta)$, 
that of $(2+N_{\rm f})$-flavor is written as
\begin{eqnarray}
V_{\rm eff}(P; \beta, m_f, \mu_f)
= V_0(P; \beta_0) - \ln R(P; \beta, m_f, \mu_f; \beta_0), \ \ 
\label{eq:vefftrans}
\end{eqnarray}
with
\begin{eqnarray}
\ln R(P; \beta, m_f, \mu_f; \beta_0)
&=& 6(\beta - \beta_0)N_{\rm site}P 
 + \ln
     \left\langle
     \displaystyle
     \left( \frac{\det M(m_{\rm l},\mu)}{\det M(m_{\rm l}   ,0)}
     \right)^{2} \prod_{f=1}^{N_{\rm f}}
     \frac{\det M(m_f,  \mu_f)}{\det M(\infty,0)}
     \right\rangle_{\! (P: {\rm fixed})}, \hspace{7mm}
\label{eq:lnr}
\end{eqnarray}
where 
$ \langle \cdots \rangle_{(P: {\rm fixed})} \equiv 
\langle \delta(P- \hat{P}) \cdots \rangle_{\beta_0} /
\langle \delta(P- \hat{P}) \rangle_{\beta_0} $
and $\langle \cdots \rangle_{\beta_0}$ means the ensemble average over
2-flavor configurations generated at $\beta_0$, $m_{\rm l}$ and
vanishing $\mu_{\rm l}$.
$\beta_0$ is the simulation point, which may differ 
from $\beta$ in this method.

At a first order transition point, $V_{\rm eff}$ shows 
a double-well shape as a function of $P$, 
and equivalently the slope of the potential $d V_{\rm eff}/dP$ shows an S-shape.
Since $\beta$ appears only in the linear term of $P$ on the right 
hand side of Eq.~(\ref{eq:lnr}), 
the shape of the slope $dV_{\rm eff}/dP$ is independent of $\beta$, 
i.e. a change of $\beta$ just shifts the overall constant of the slope of $V_{\rm eff}$ \cite{eji07}.
Although $\beta$ must be adjusted to the first order transition point to observe the double-well potential, the fine tuning is not necessary if we investigate the slope.

The derivative $dV_0 /dP$ can be measured easily from the peak position of the plaquette histogram \cite{yoneyama09}.
When one performs a simulation at $\beta_0$, the slope is vanishing at the
minimum of $V_0(P;\beta_0)$, and the value of $P$ at the minimum
can be estimated by $\langle P \rangle_{\beta_0}$ approximately. 
Hence, we obtain $dV_0/dP$ at $\langle P \rangle_{\beta_0}$ by
\begin{equation}
\frac{dV_0}{dP} ( \langle P \rangle_{\beta_0}, \beta) 
= - 6(\beta - \beta_0)N_{\rm site}.
\label{eq:dveffdp}
\end{equation}
We therefore focus on the slope of the effective potential to identify the nature of transitions.

\section{Light quark mass dependence of the critical heavy quark mass}
\label{sec:heavy}

\begin{figure}[tb]
\begin{center}
\vspace{-5mm}
\centerline{
\includegraphics[width=77mm]{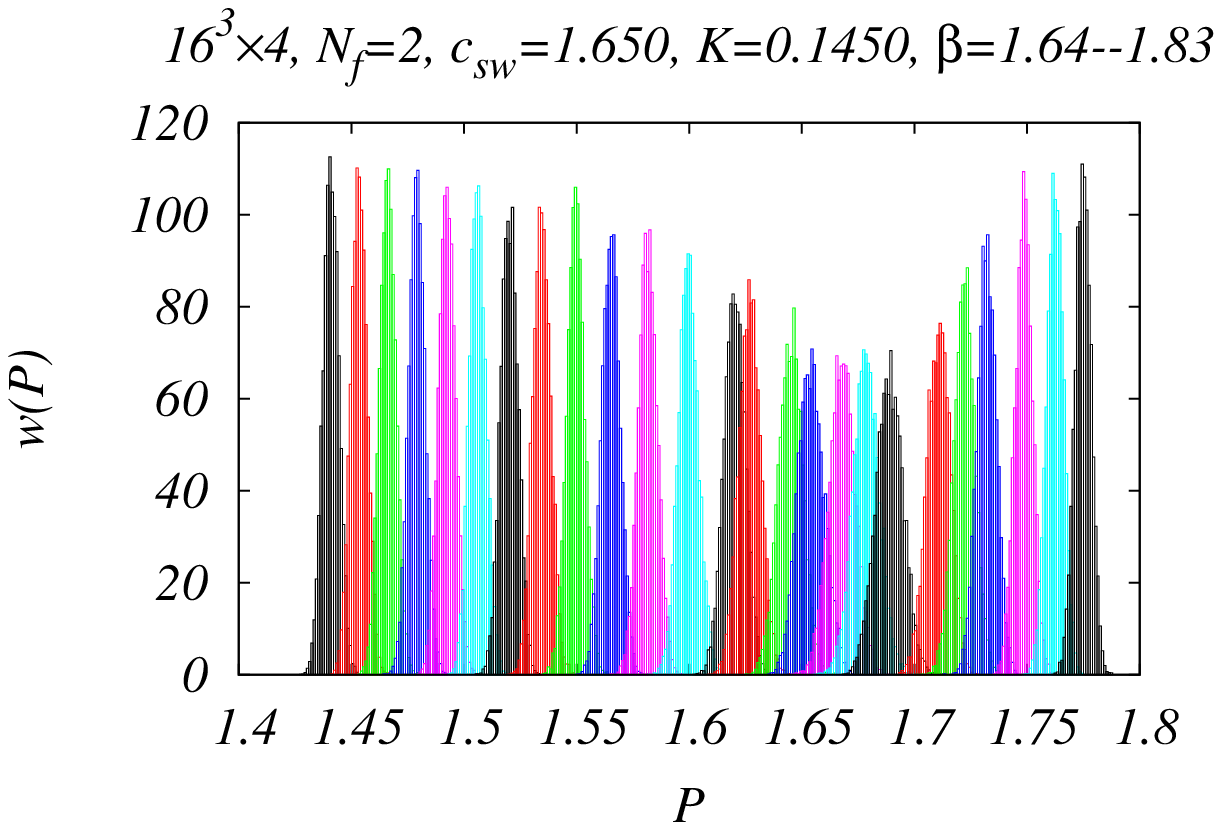}
\includegraphics[width=77mm]{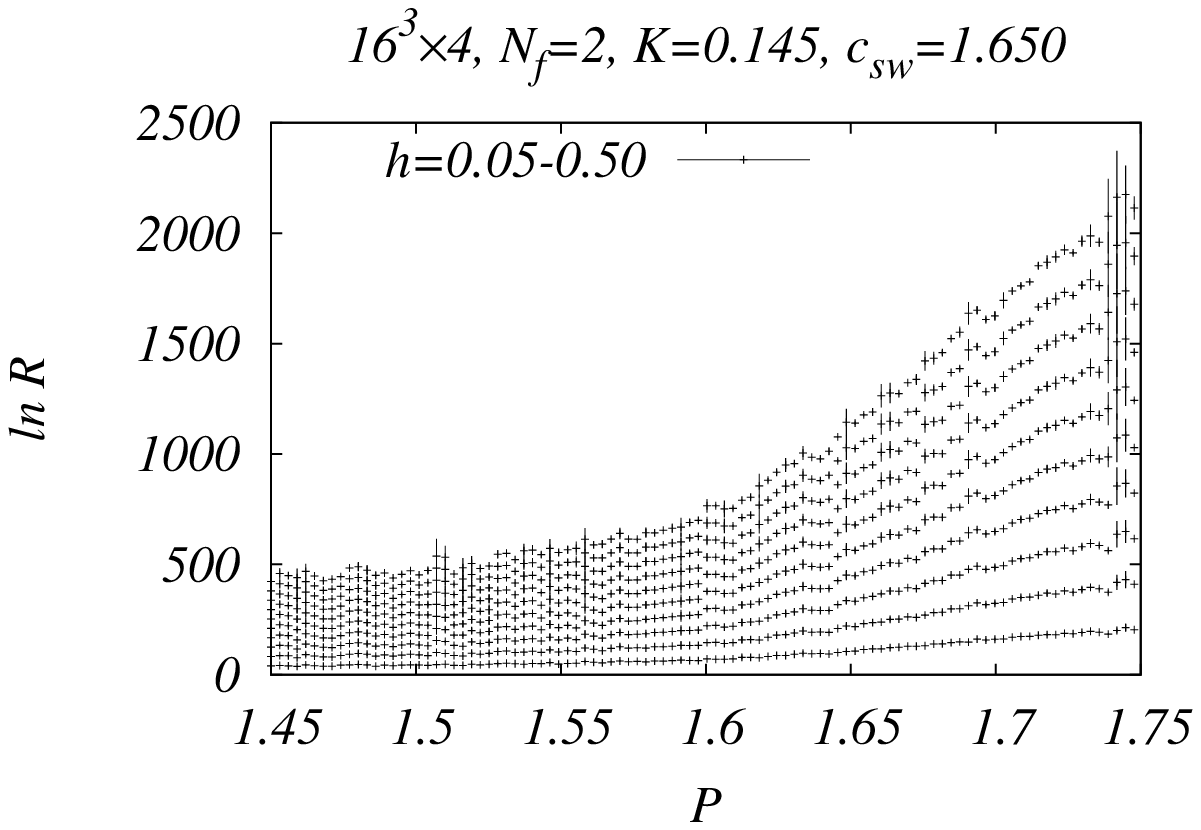}
}
\vspace{0mm}
\caption{Left: Plaquette histogram of 2-flavor QCD $w_0 (P; \beta, \kappa,0)$ at $\kappa=0.145$. 
Right: $\ln \bar{R} (P; h,0)$ as a function of $P$ for $h=0.05$ -- $0.50$.}
\label{fig1}
\end{center}
\end{figure}

\begin{figure}[tb]
\begin{center}
\vspace{-5mm}
\centerline{
\includegraphics[width=75mm]{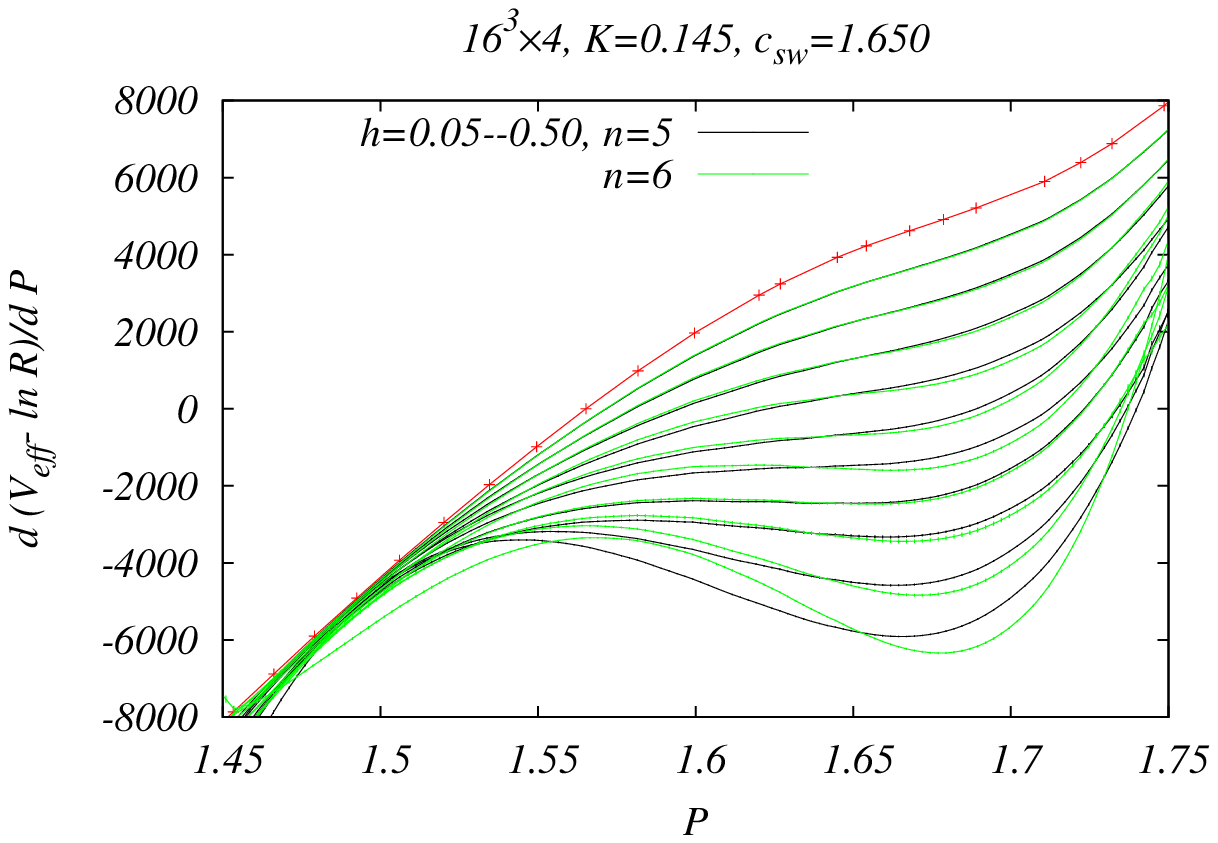}
\includegraphics[width=75mm]{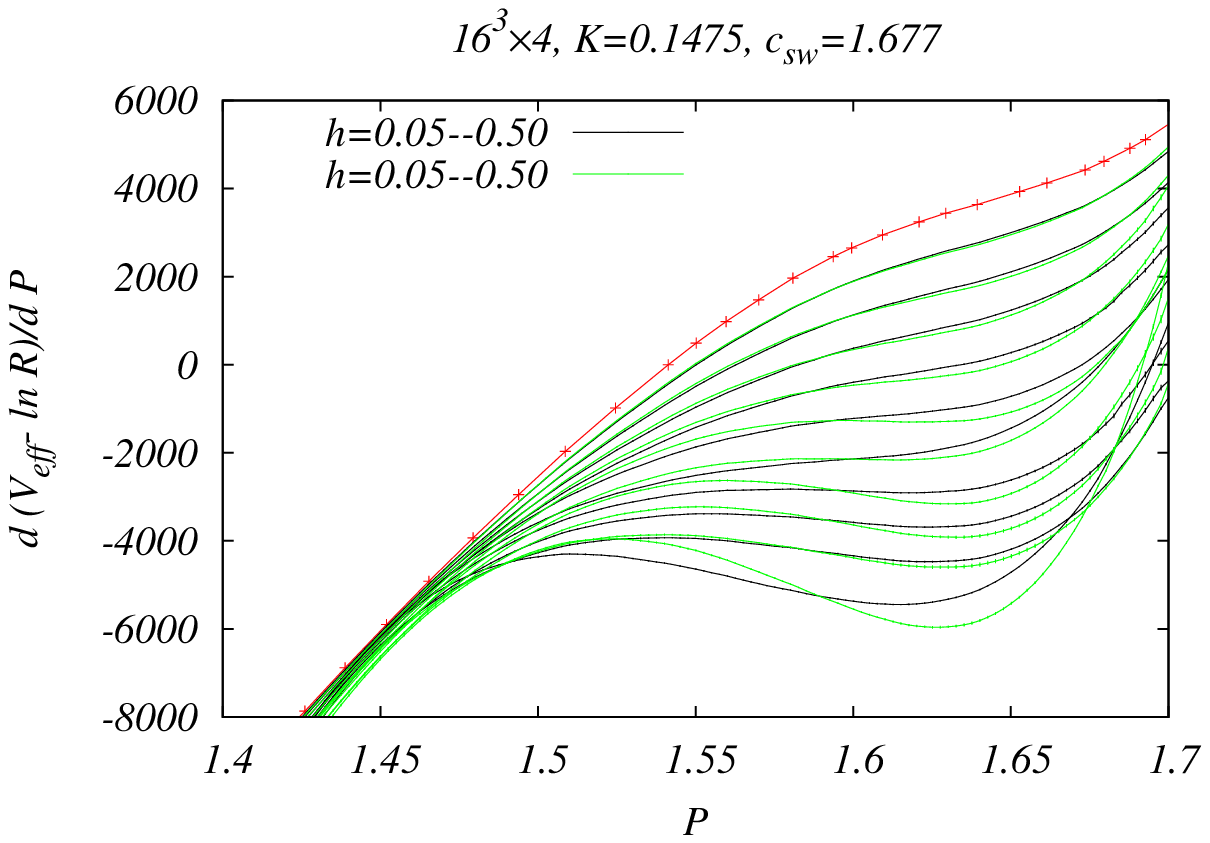}
}
\centerline{
\includegraphics[width=75mm]{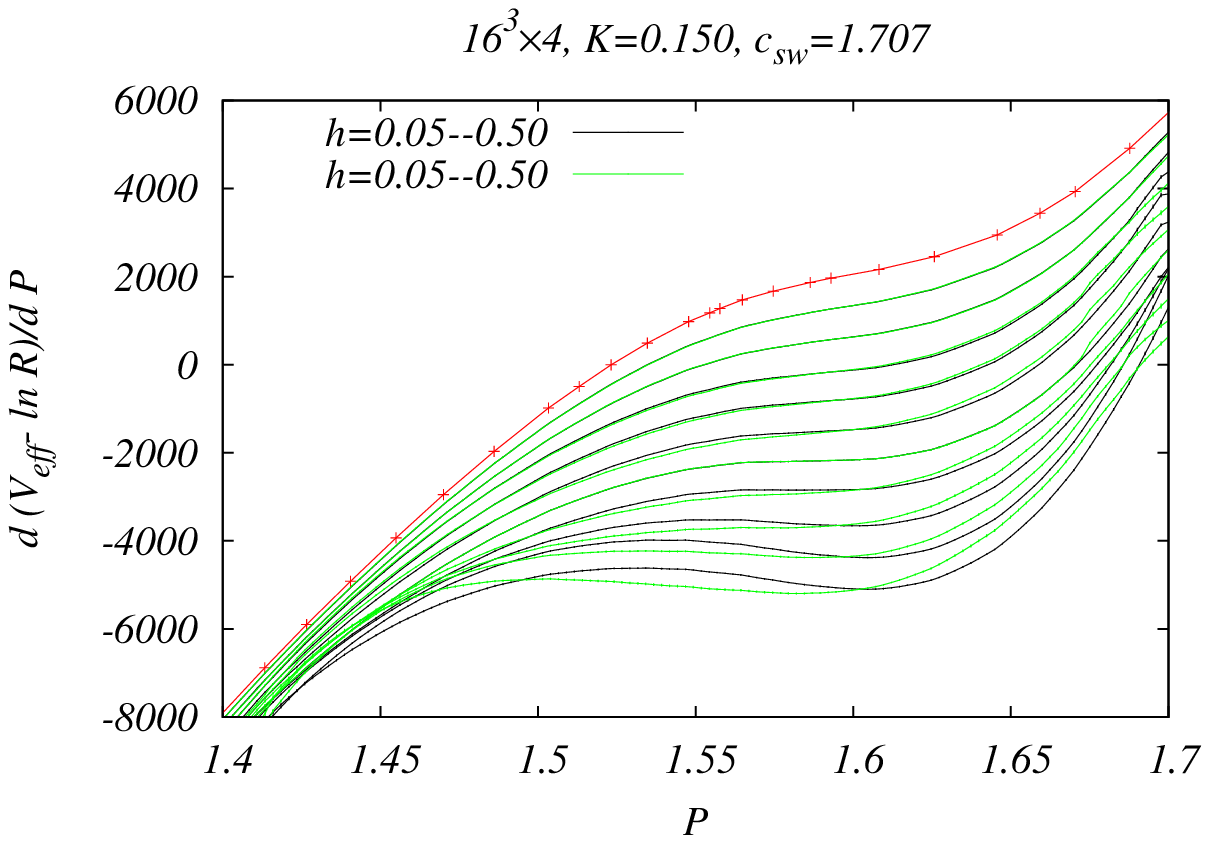}
\includegraphics[width=75mm]{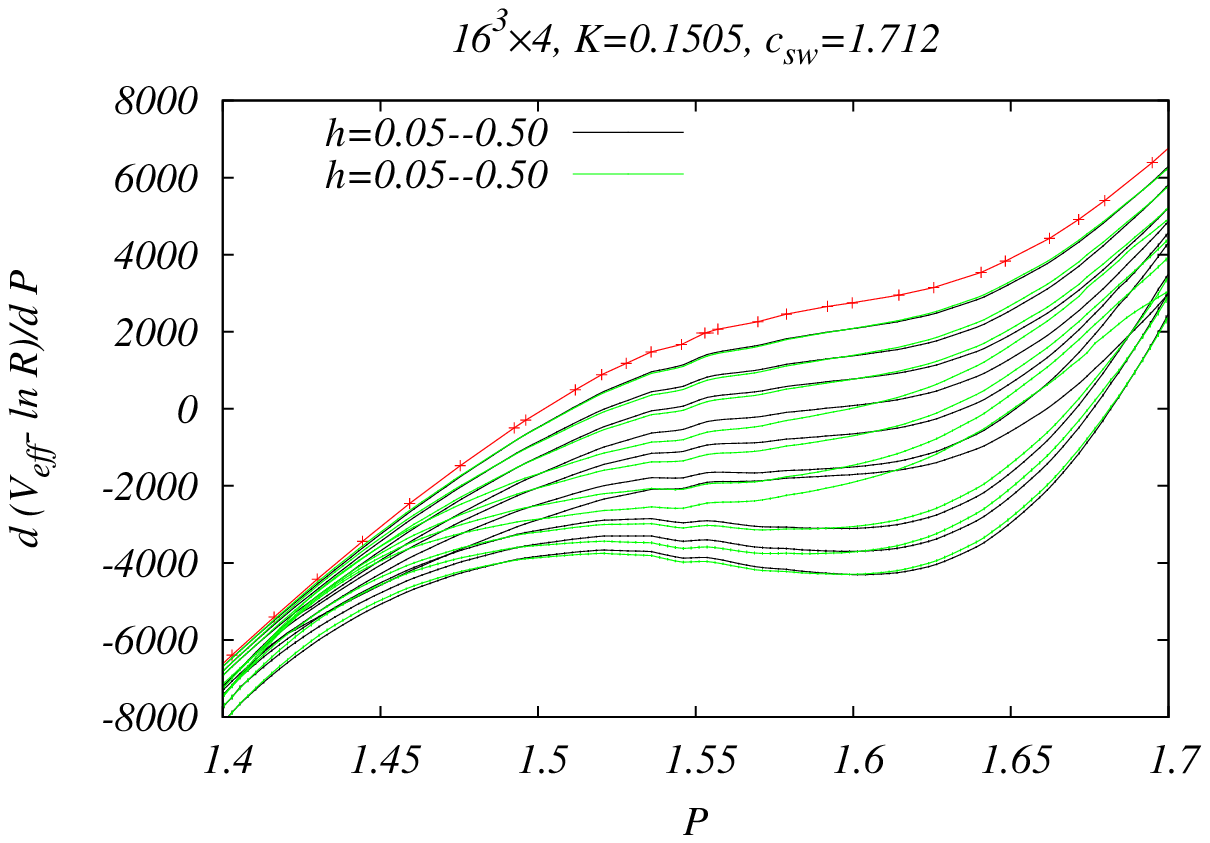}
}
\vspace{0mm}
\caption{$dV_{\rm eff}/dP(P; \beta, \kappa, h)$ as a function of $P$ at $\kappa_l=0.1450$(top left), $0.1475$ (top right), $0.1500$ (bottom left) and $0.1505$ (bottom right). 
}
\label{fig2}
\end{center}
\end{figure}

\begin{figure}[tb]
\vspace{-3mm}
\begin{center}
\centerline{
\includegraphics[width=75mm]{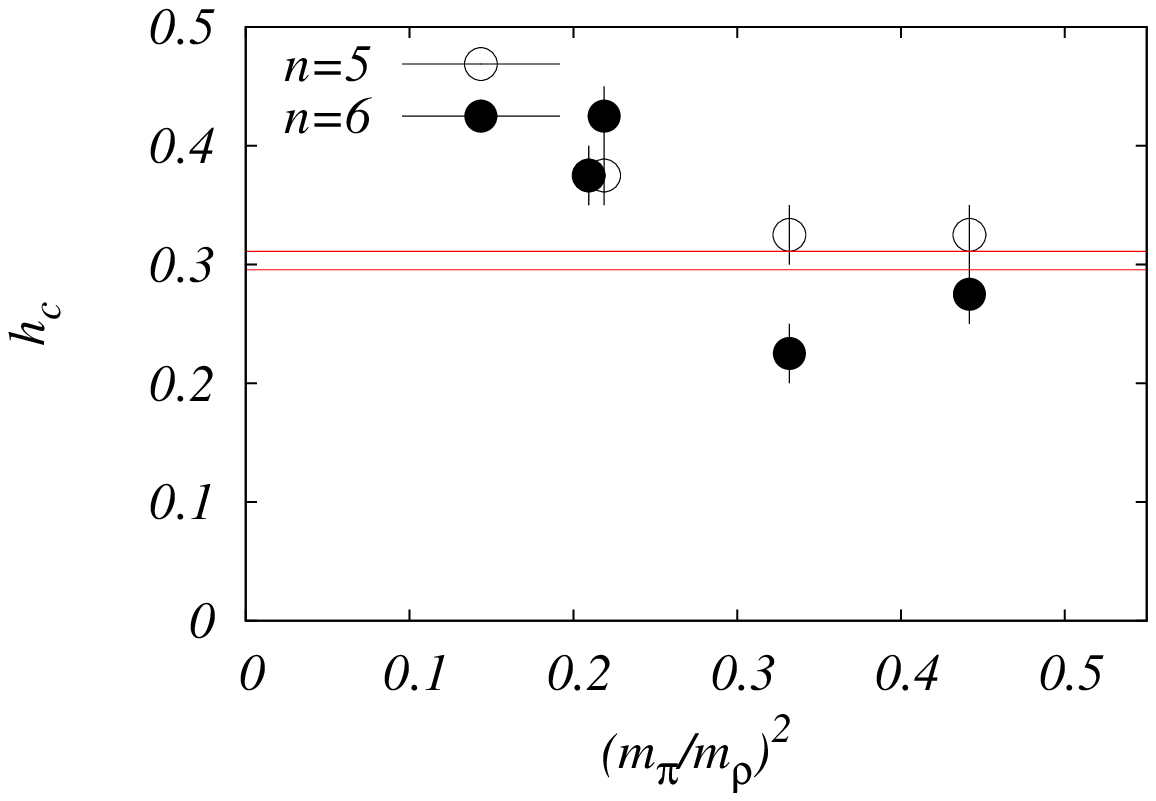}
\includegraphics[width=75mm]{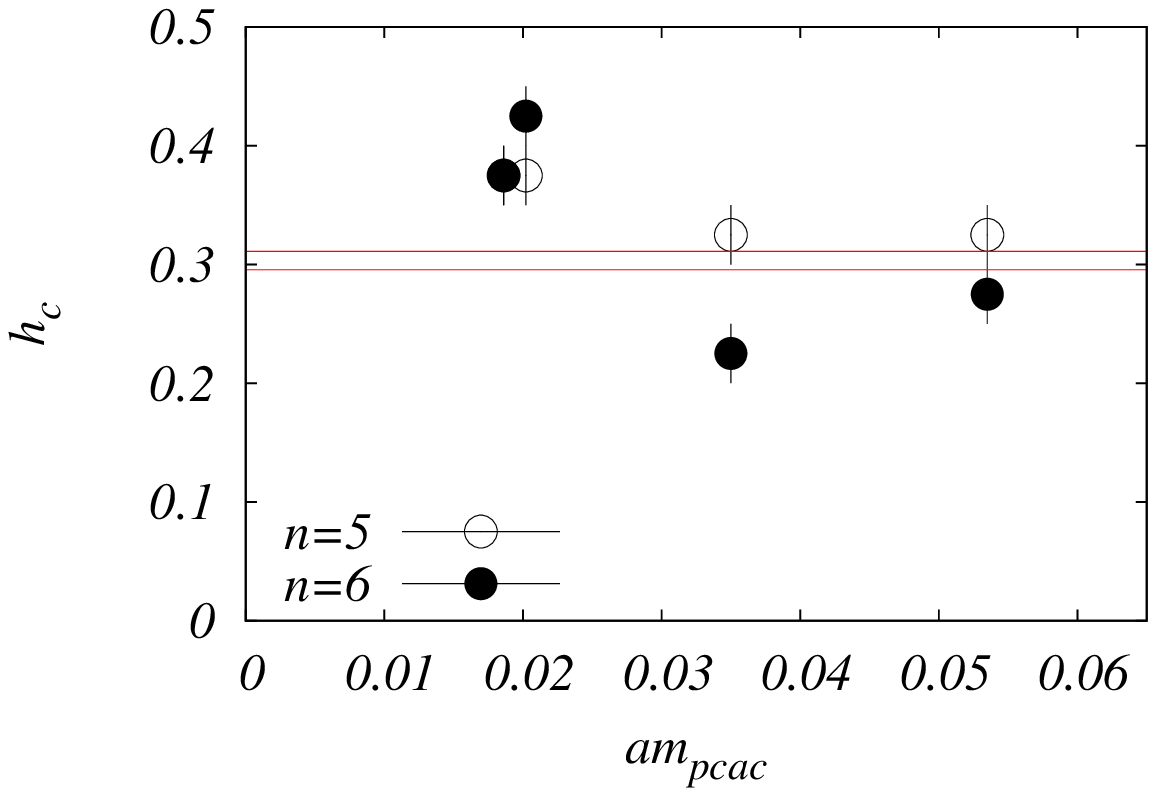}
}
\caption{The critical $h$ as a function of $(m_{\pi}/m_{\rho})^2$ (left) and of $m_{\rm PCAC}$ (right).
The red liens are lower and upper bounds of the critical value of $h$ obtained by dynamical simulations of degenerate 16-flavor QCD. 
}
\label{fig3}
\end{center}
\end{figure}

In this section, we discuss the light quark mass dependence of the critical hopping parameter of heavy quarks terminating the first order transition.
It is important to investigate whether the critical hopping parameter vanishes at a finite light quark mass or not.
If the critical line crosses the line of $\kappa_{\rm h}=0$, the 2-flavor chiral limit is in the first order region, where $\kappa_{\rm h}$ is the hopping parameter of heavy-flavors being proportional to the inverse mass. 
Restricting the calculation to the heavy quark region near $\kappa_{\rm h}=0$,
the second determinant for $N_{\rm f}$ flavors in Eq.~(\ref{eq:lnr}) is approximated
by the leading order of the hopping parameter expansion,
\begin{eqnarray}
\ln \left[ \frac{\det M (\kappa_{\rm h}, \mu_{\rm h})}{\det M (0,0)} \right]  
=  288 N_{\rm site} \kappa_{\rm h}^4 \hat{P} + 12 N_s^3 (2 \kappa_{\rm h})^{N_t} 
\left( \cosh(\mu_{\rm h} /T) \hat{\Omega}_{\rm R} 
+i \sinh(\mu_{\rm h} /T) \hat{\Omega}_{\rm I} \right) + \cdots \ \ 
\label{eq:detmw}
\end{eqnarray}
for the standard Wilson quark action \cite{whot14}.
$\hat{\Omega}_{\rm R}$ and $\hat{\Omega}_{\rm I}$ are the real and imaginary 
part of the Polyakov loop, respectively.
For improved gauge actions such as 
$S_g = -6N_{\rm site} \beta [c_0 {\rm (plaquette)} 
+ c_1 {\rm (rectangle)}]$, 
additional $c_1\times O(\kappa^4)$ terms must be considered, 
where $c_1$ is the improvement coefficient and $c_0=1-8c_1$. 
However, since the improvement term does not affect the physics, 
we will cancel these terms by a shift of the coefficient $c_1$.
It is shown in Ref.~\cite{yamada13} that the hopping parameter expansion is applicable for large $N_{\rm f}$.

Denoting $h=2 N_{\rm f} (2\kappa_{\rm h})^{N_t}$, 
we obtain \ 
$\ln R(P;\beta,\kappa_{\rm h},0;\beta_0)
=\ln\bar{R}(P; h,0)
+({\rm plaquette \ term})$ 
$+O(\kappa_{\rm h}^{N_t+2})$
for $\mu_{\rm l}=\mu_{\rm h}=0$ with
\begin{eqnarray}
\bar{R}(P; h,0)
= \left\langle \exp [6h N_s^3 \hat{\Omega}_{\rm R}] 
\right\rangle_{(P: {\rm fixed}, \beta_0)} ,
\label{eq:rew2f}
\end{eqnarray}
where $\bar{R}(P; h,0)$ is given by the Polyakov loop and is independent of $\beta_0$.
The plaquette term does not contribute to $d^2V_{\rm eff}/dP^2$
and can be absorbed by shifting  
$\beta \to \beta^{*} \equiv \beta + 48 N_{\rm f} \kappa_{\rm h}^4$ for Wilson quarks. 
Moreover, one can deal with the case with non-degenerate masses by adopting
$h=2 \sum_{f=1}^{N_{\rm f}} (2 \kappa_f)^{N_t}$ for the Wilson quark action or 
$h=(1/4)\sum_{f=1}^{N_{\rm f}} (2m_f)^{-N_t}$ for the staggered quark action.
Thus, the choice of the quark action is not important.
In the following, we discuss the mass dependence of $\bar{R}$ 
through the parameter $h$.

We perform simulations of 2-flavor QCD with the clover-improved
Wilson quark and Iwasaki gauge actions\footnote{The improvement parameters are almost the same as Ref.~\cite{whot09}}.
We take four different values of light quark masses ranging from $\kappa_l = 0.145$ to $0.1505$.
The corresponding ratio of pseudo-scalar and vector meson masses is
$m_{\rm PS}/m_{\rm V} \approx 0.6647,$ $0.5761,$ $0.4677$ and $0.4575$ 
for $\kappa_l =0.145,$ $0.1475,$ $0.150$ and $0.1505$, respectively.
The data are taken at about 30 values of $\beta$ around the pseudo-critical point at each $\kappa_l$, and at each simulation point 10,000 to 40,000 trajectories have been accumulated.

In the calculation of $\bar{R}(P; h,0)$, we use the delta
function approximated by
$\delta(x) \approx 1/(\Delta \sqrt{\pi})$ $\exp[-(x/\Delta)^2]$, 
where $\Delta$ is selected consulting the resolution and the statistical error. 
Because $\bar{R}(P; h,0)$ is independent of $\beta$, we mix all
data obtained at different $\beta$ as was done in Ref.~\cite{eji07}.
The histograms of plaquette value are plotted in the left panel of Fig.~\ref{fig1}.
The results for $\ln \bar R(P;h,0)$ at $\kappa_{\rm l}=0.145$ are shown in the right panel of Fig.~\ref{fig1} for $h=0.05$ -- $0.50$ at interval of $0.05$. 
A rapid increase is observed around $P \sim 1.6$, and the gradient becomes larger as $h$ increases.

The derivative $d \ln \bar R/dP$ is calculated by fitting $\ln \bar R$ 
to a $n^{\rm th}$-order polynomial of $P$ in an appropriate fit range.
Using the equation,
\begin{eqnarray}
\frac{d V_{\rm eff}}{dP} = \frac{d V_0}{dP} - \frac{d (\ln \bar R)}{dP} + {\rm (const.)}, 
\end{eqnarray}
we compute $dV_{\rm eff}/dP$ for each $\kappa_l$. 
The results for various $h$ are shown in Fig.~\ref{fig2}, where 
$dV_0/dP$ is computed by Eq.~(\ref{eq:dveffdp}).
The shape of $dV_{\rm eff}/dP$ is independent of $\beta$ because 
$d^2 V_{\rm eff}/dP^2$ is $\beta$-independent.
The first derivative $dV_{\rm eff}/dP$ is the monotonically increasing function of $P$ when $h$ is small, indicating that the transition is crossover.
However, the shape of $dV_{\rm eff}/dP$ turns into an S-shaped function
at $h \approx 0.3$, which means that the system undergoes first order transition.
The same analysis has been done in Ref.~\cite{yamada13} using the p4-improved staggered fermion action for 2-flavor QCD with $m_{\rm PS}/m_{\rm V} \approx 0.7$. 
The result of the critical value of $h$ at which the first order transition appears, $h_c$, is about 0.06. 
The difference may be caused by the lattice discretization error due to small $N_t$.
We have defined the parameter $h=2 N_{\rm f} (2\kappa_{\rm h})^{N_t}$ for the
Wilson quark.
Then, the critical point $\kappa_{hc}$ corresponding $h_c$ decreases as
$\kappa_{hc} \propto (h_c/N_{\rm f})^{1/N_t}$
with $N_{\rm f}$, and 
the truncation error from the higher order terms of the hopping parameter expansion 
in $\kappa_{\rm h}$ becomes smaller as $N_{\rm f}$ increases.

The remarkable point of this study is light quark mass dependence. 
In Fig.~\ref{fig3}, we plot the results of the critical value $h_c$ as functions of 
$(m_{\rm PS}/ m_{\rm V})^2$ (left) and the quark mass defined by the PCAC relation $m_{\rm pcac}$ (right). The slope of $V_{\rm eff}$ is computed fitting the data with $5^{\rm th}$ or $6^{\rm th}$ order polynomials. The open symbols are the results by $5^{\rm th}$ order, and the filled symbols are those by $6^{\rm th}$ order.
The difference is taken as the systematic error.
The red lines are the upper and lower bounds of the critical $h$ for the system with $N_{\rm f}=16$ massive quarks and no light quarks, which is determined by observing the hysteresis curves of the plaquette and Polyakov loop.
It is found from these figures that the light quark mass dependence of the critical mass of heavy quarks is very small in the region we computed. 
The weak dependence suggests that the critical mass of heavy quark remains finite in the chiral limit of 2-flavors. Thus, the sign of the first order transition in 2-flavor QCD is not shown.

\section{Singularities in the complex chemical potential plane}
\label{sec:singularity}

Let us turn to finite density QCD.
The distribution function of 2-flavor QCD at finite density and the appearance of the double-well potential are discussed in Ref.~\cite{eji07}, and the boundary of the first order transition region of $(2+N_{\rm f})$-flavor QCD at finite density is computed in Ref.~\cite{yamada13} using the 2-flavor QCD configurations generated with the p4-improved staggered quark action in Ref.~\cite{BS05}.
The first order region is found to become larger as $\mu$ increases.

In this section, we extend the potential analysis to the complex chemical potential,
$\mu= \mu_{\rm Re} + i \mu_{\rm Im}$.
The numerical study of the singularities of ${\cal Z}=0$ has a potential danger for large $\mu$ when we use the reweighting method \cite{eji05}.
We estimate the position of ${\cal Z}=0$ from the probability distribution of the complex phase, since ${\cal Z}$ vanishes when the distribution function has two peaks or more and the contributions from these peaks cancel each other \cite{eji05}.
In this study, we investigate the plaquette distribution function and complex phase average with constraining the plaquette value.
To avoid the sign problem, the Gaussian approximation \cite{eji07} is applied 
\footnote{Preliminary results are presented in Ref.~\cite{yoneyama09}}.

We compute the probability distribution function of the plaquette $P$, 
\begin{eqnarray}
w(P,\beta,\mu) = 
\int {\cal D} U \ \delta(P- \hat{P}) \ (\det M)^2 
e^{6\beta N_{\rm site} \hat{P}}. 
\label{eq:pdistfd}
\end{eqnarray}
We denote $ w_0(P,\beta) \equiv w(P,\beta,0)$.
The normalized partition function is rewritten as 
\begin{eqnarray}
\frac{{\cal Z}(\beta, \mu)}{{\cal Z}(\beta, 0)} 
=  \frac{1}{{\cal Z} (\beta,0)} \int w(P,\beta,\mu) \ dP
= \frac{1}{{\cal Z} (\beta,0)} \int R(P,\mu) w_0 (P,\beta) \ dP.
\label{eq:rewmu}
\end{eqnarray}
Here, $R(P,\mu)$ is the reweighting factor for finite $\mu$ 
defined by Eq.~(\ref{eq:lnr}) with $\beta_0=\beta$ and $m_f= \infty$ for $N_{\rm f}$-flavors.
This $R(P, \mu)$ is independent of $\beta$ and $R(P, \mu)$ can be measured at any $\beta$. 
In this study, all simulations are performed at $\mu=0$ and the effect of 
finite $\mu$ is introduced though the operator 
$\det M(\mu) / \det M(0)$ measured on the configurations 
generated by the simulations at $\mu=0$.

Because QCD has time-reflection symmetry, the partition function is invariant 
under a change from $\mu$ to $-\mu$, i.e. $R(P,-\mu) = R(P,\mu)$.
Moreover, the quark determinant satisfies $\det M (-\mu) = (\det M(\mu^*))^*$.
From these equations, we get 
\begin{eqnarray}
[R(P, \mu)]^* = R(P, \mu^*).
\end{eqnarray}
This indicates that $R(P, \mu)$ is real valued function in the case of real $\mu$, 
i.e. $\mu = \mu^*$.
Then, the plaquette probability distribution 
$R(P, \mu) w_0(P, \beta)$ is also real valued. 
However, once the imaginary part of $\mu$ becomes nonzero, 
$R(P, \mu)$ is not a real number any more.
We thus write the partition function,
\begin{eqnarray}
{\cal Z}(\beta, \mu) 
= \int e^{i \phi(P,\mu)} |R(P,\mu)| w_0(P,\beta) \ dP .
\label{eq:phidis}
\end{eqnarray}
This complex phase $\phi$ vanishes at $\mu_{\rm Im}=0$, and 
$\phi$ is monotonic function of $\mu_{\rm Im}$ at small $\mu$ if we define the complex phase by a Taylor expansion of $\ln \det M (\mu)$.

If $|R(P,\mu)| w_0 (P,\beta)$ becomes a double-peaked function which has two peaks of equal height at $P_+$ and $P_-$, two phases coexist like first order phase transitions.
When changing $\mu$, the expectation value of plaquette changes discontinuously form 
$P_- e^{i \phi(P_-)}$ to $P_+ e^{i \phi(P_+)}$ at that point.
And then, the partition function can be approximated by 
$Z \approx (e^{i \phi(P_+)} +e^{i \phi(P_-)}) \times {\rm (const.)}$.
When the difference between $\phi(P_+)$ and $\phi(P_-)$ is equal to $(2n+1)\pi$ with an integer $n$, the partition function will be vanishing, i.e. Lee-Yang zeros appear.
The first-order-transition-like point in the complex plane corresponds to ``Stokes line'' in the infinite volume limit.
We define the effective potential as 
$V_{\rm eff} = \ln [|R(P, \mu)| w_0(P, \beta)]$ 
and investigate the position of this would-be Stokes line in the complex $\mu$ plane.

We calculate these three quantities, $|R(P,\mu)|$, $\phi (P,\mu)$ and $w_0 (P,\beta)$ by Monte-Carlo simulations. 
Because the exact calculation of the quark determinant is difficult except on small lattices, we estimate the quark determinant from the data of Taylor expansion coefficients of $\ln \det M (\mu)$ up to $O(\mu^6)$ around $\mu=0$ obtained by a simulation of 2-flavor QCD with p4-improved staggered quarks in Ref.~\cite{BS05}.
The truncation error has been discussed in Ref.~\cite{eji07}.
Notice that the complex phase $\theta$ is not restricted to the range from $-\pi$ to $\pi$ because we define $\theta$ by the Taylor expansion of $\ln \det M$.

To compute $V_{\rm eff}$ at finite $\mu$, we discuss the sign problem.
We denote the quark determinant as
$N_{\rm f} \ln[\det M(\mu) / \det M(0)] \equiv F +i \theta$.
Histograms of $\theta$ seem to be well-approximated by Gaussian functions.
(See Fig.~1 (left) in Ref.~\cite{yoneyama09}.)
Here, we perform a cumulant expansion, 
\begin{eqnarray}
\langle \exp(F +i \theta) \rangle = \langle e^F \rangle 
\exp \left[ i\langle \theta \rangle 
- \frac{1}{2} \langle (\Delta \theta)^2 \rangle
- \frac{i}{3!} \langle (\Delta \theta)^3 \rangle 
+ i \langle \Delta F \Delta \theta \rangle 
- \frac{1}{2} \langle \Delta F (\Delta \theta)^2 \rangle + \cdots \right]
\label{eq:cum}
\end{eqnarray}
with $\Delta X= X - \langle X \rangle$. 

For the case that the distribution of $\theta$ is of Gaussian, the $O(\theta^n)$ terms vanish for $n >2$ in this equation. 
Moreover, since $\theta \sim O(\mu)$ and $F \sim O(\mu^2)$, this expansion can be regarded as a power expansion in $\mu$. 
If the convergence of the expansion Eq.~(\ref{eq:cum}) is good, we can extract the phase factor $e^{i \phi}$ from $R$ easily and the sign problem in $|R|$ is eliminated.
We deal with the first two terms, 
i.e. $i \langle \theta \rangle$ and 
$-\langle (\Delta \theta)^2 \rangle /2$,
assuming the Gaussian distribution. 
The correlation terms between $F$ and $\theta$ are also neglected as a first step.
Because we calculate the expectation value with fixed $P$ and the values of $F$ and $P$ are strongly correlated, the $\Delta F$ may be small once $P$ is fixed.
Then, $\phi \approx \langle \theta \rangle$.

\begin{figure}[t]
\vspace{-5mm}
\begin{center}
\includegraphics[width=70mm]{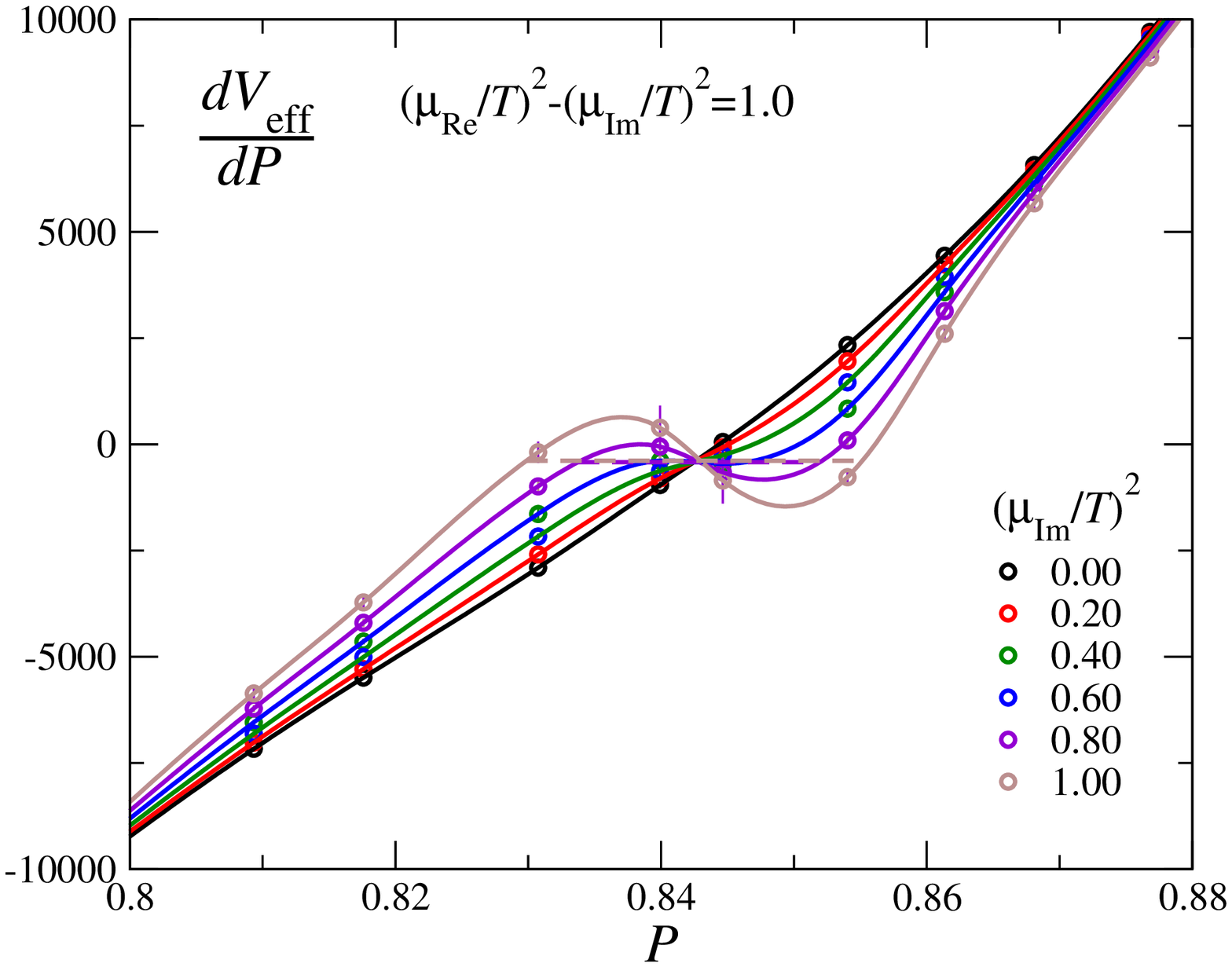}
\includegraphics[width=70mm]{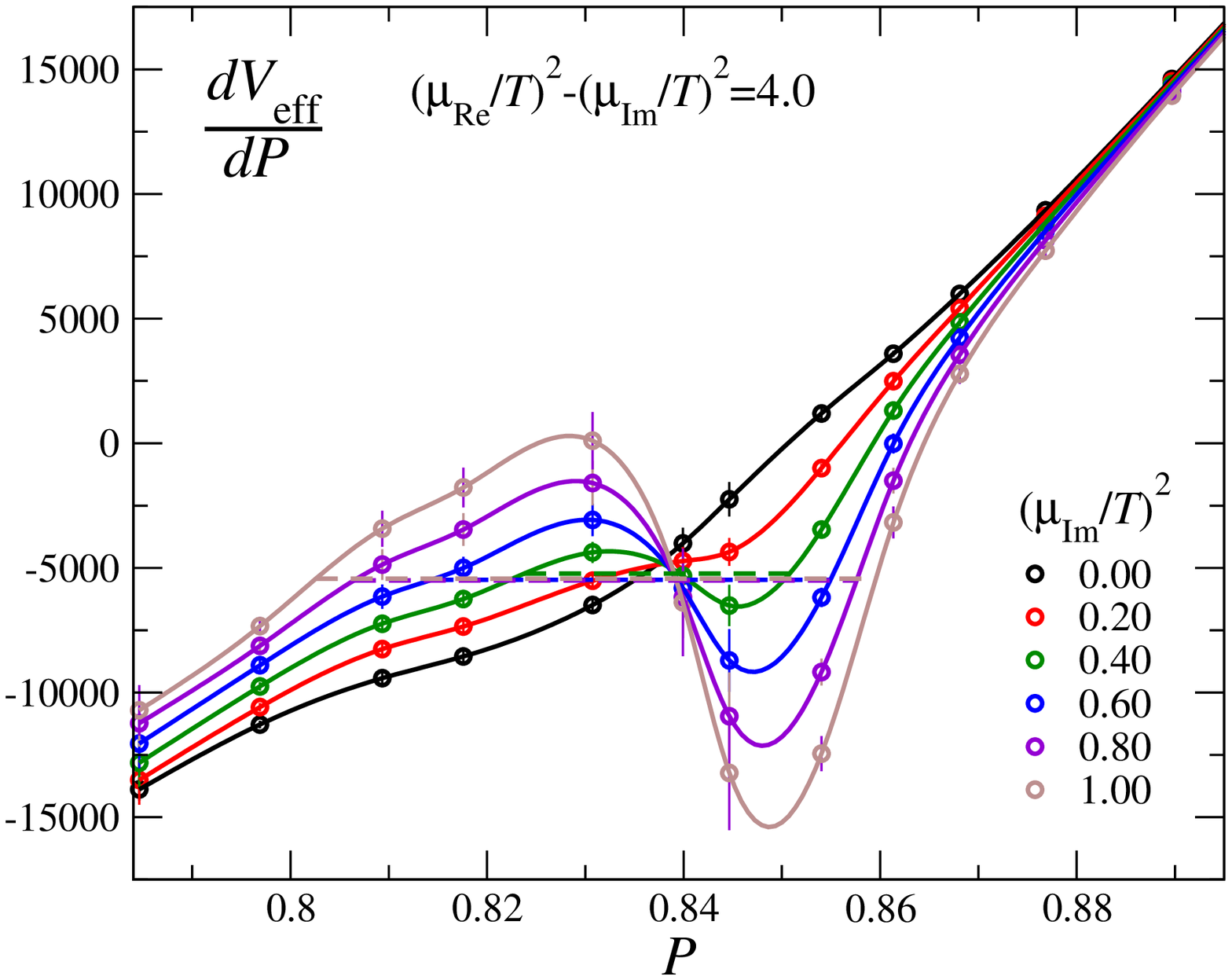}
\caption{
The derivative of the effective potential $dV_{\rm eff}/dP$ at $\beta=3.65$ 
for $(\mu_{\rm Re}/T)^2 - (\mu_{\rm Im}/T)^2 =1.0$ (left) and $4.0$ (right). 
}
\label{fig4}
\end{center}
\vskip -0.3cm
\end{figure} 

\begin{figure}[tb]
\vspace{-5mm}
\begin{center}
\centerline{
\includegraphics[width=70mm]{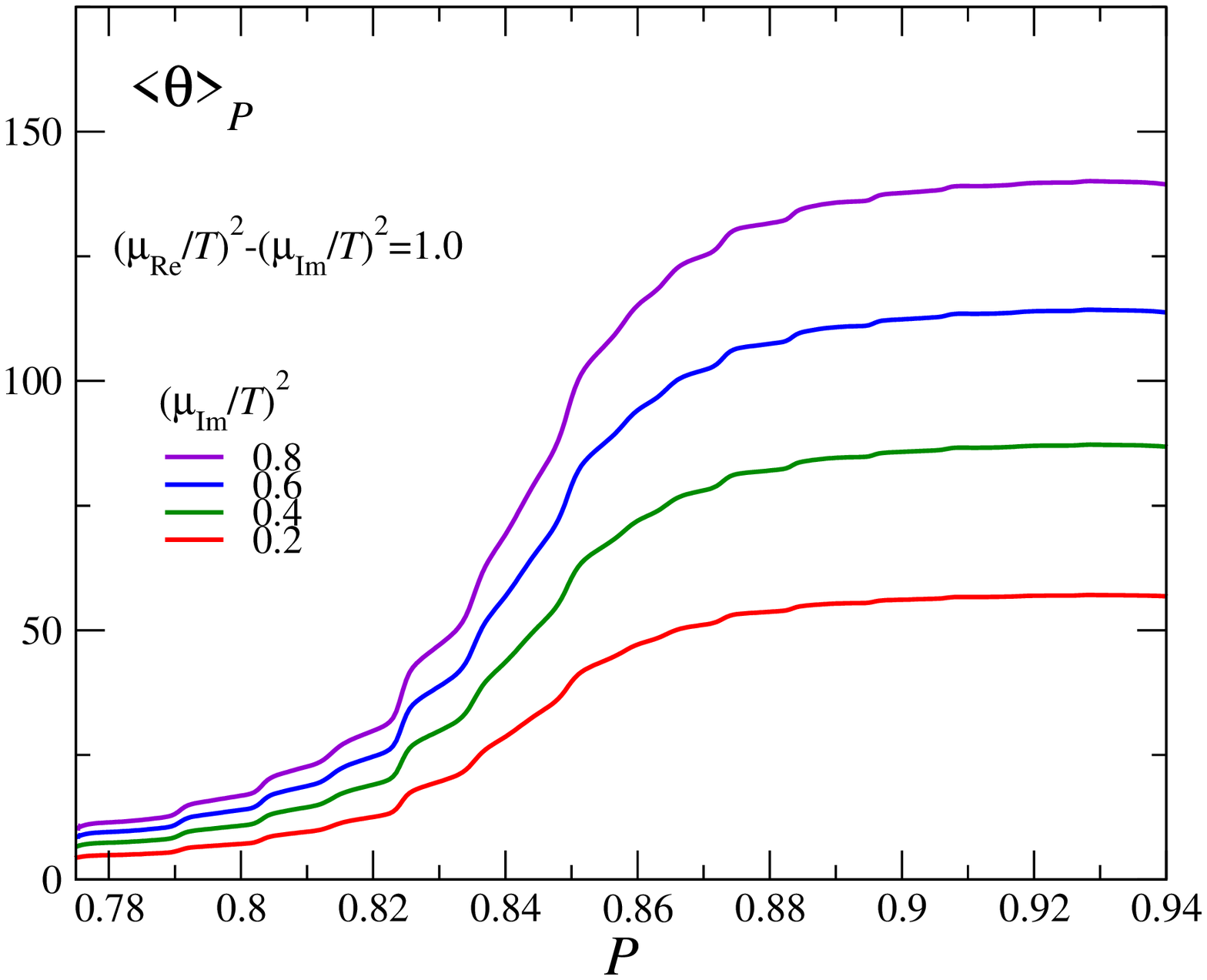}
\includegraphics[width=70mm]{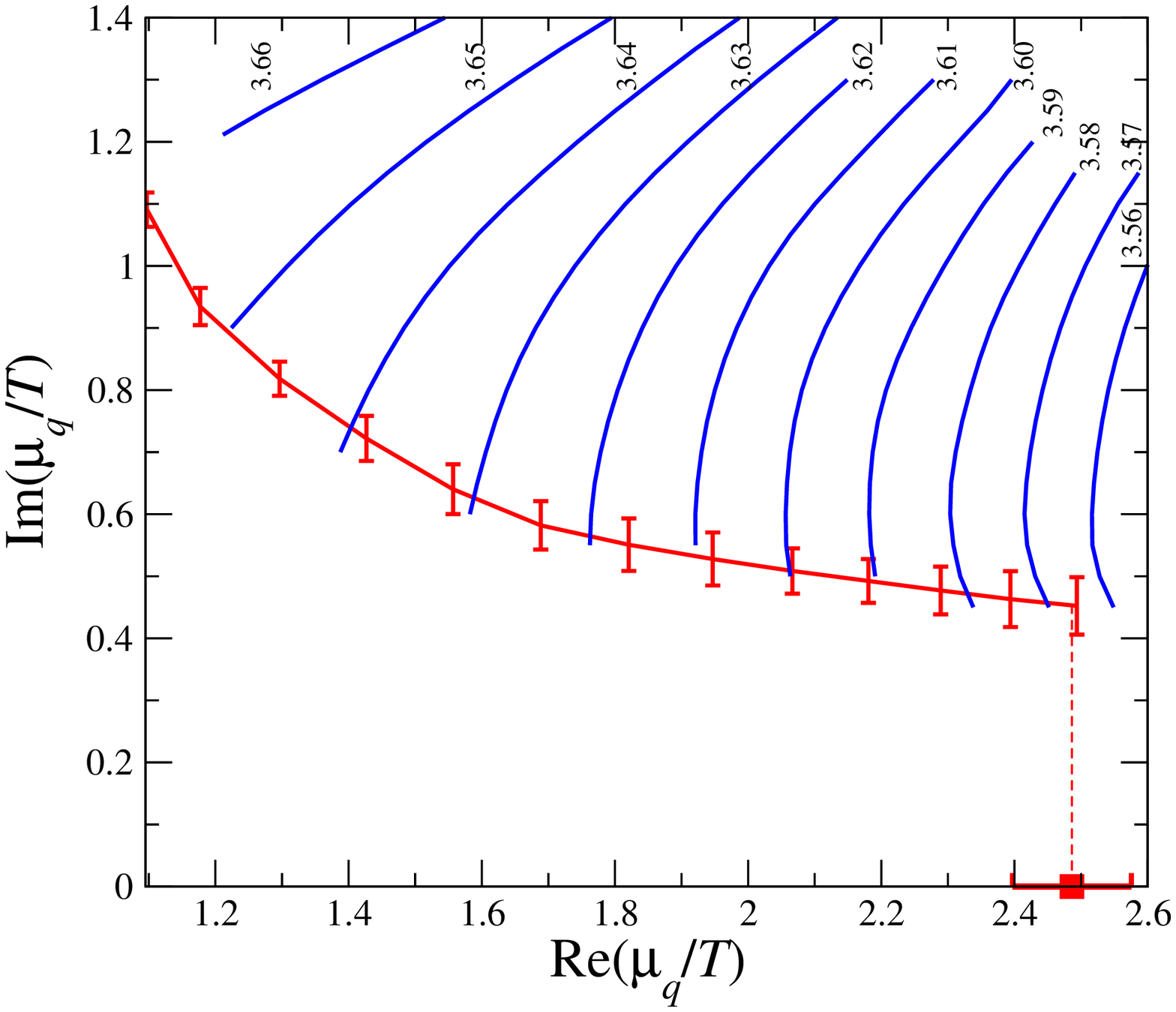}
}
\caption{
Left: The complex phase $\phi$ for ${\rm Re}(\mu/T)^2=1$ as a function of $P$.
Right: The blue lines are the would-be Stokes lines for each $\beta$.
The values of $\beta$ are shown near the blue lines.
The red curve is the boundary where $V_{\rm eff}$ changes to double-well type 
in the complex $\mu/T$ plane.
Below this line, $V_{\rm eff}$ is always of single-well.
}
\label{fig5}
\end{center}
\end{figure}

We discuss $d V_{\rm eff}/dP$ instead of $V_{\rm eff}$ itself, again. 
$d V_{\rm eff}/dP$ at different $\beta$ can be estimated by the equation,  
\begin{eqnarray}
\frac{d V_{\rm eff}}{dP} (P,\beta)
= \frac{d V_{\rm eff}}{dP} (P,\beta_0) -6 (\beta - \beta_0) N_{\rm site},
\label{eq:derrewbeta}
\end{eqnarray}
under the parameter change from $\beta_0$ to $\beta$. 

If the effective potential $V_{\rm eff}$ is a double-well function of $P$ having two minima at $P_+$ and $P_-$ and one maximum at the middle $P_0$, 
$d V_{\rm eff}/dP$ is an S-shaped function and vanishes three times at $P_+$, $P_-$ and $P_0$.
The condition, 
\begin{eqnarray}
\int_{P_-}^{P_0} (d V_{\rm eff}/dP) dP 
=- \int_{P_0}^{P_+} (d V_{\rm eff}/dP) dP, 
\label{eq:maxcon}
\end{eqnarray}
is satisfied when $V_{\rm eff}(P_+) =V_{\rm eff}(P_-)$ at the transition point. 
In such a case, there exists a region of $P$ where the derivative of $ d V_{\rm eff}/d P$ is negative. 
The results of the first derivative of $V_{\rm eff}(P,\beta,\mu)$ are shown in Fig.~\ref{fig4} for various $\mu/T$ with ${\rm Re} [(\mu/T)^2]= (\mu_{\rm Re}/T)^2-(\mu_{\rm Im}/T)^2=1$ (left) and $4$ (right), where $\beta=3.65$.
We measured them at the peaks of the plaquette histograms and interpolated the data by a cubic spline method.
In the region of large $\mu_{\rm Im}/T$, $d V_{\rm eff}/d P$ becomes an S-shaped function. 
The result of $\phi$ is plotted in Fig.~\ref{fig5} (left) for ${\rm Re}[(\mu/T)^2] = 1$. It is a monotonically increasing function of $P$.

We then investigate the boundary at which $dV_{\rm eff}/d P$ changes to an S-shaped function from a monotonic function. 
The boundary is shown as a red line in Fig.~\ref{fig5} (right).
Above this line, the region of $P$ where $d^2 V_{\rm eff}/dP^2=0$ appears. 
The constant part of $dV_{\rm eff}/d P$ is changed by varying $\beta$.
We then adjust $\beta$ such that the depths of two minima of $V_{\rm eff}$ are the same using Eq.~(\ref{eq:derrewbeta}).
If the dashed lines in Fig.~\ref{fig4} move to the horizontal axis by changing $\beta$, Eq.~(\ref{eq:maxcon}) is satisfied.

Contour plots of the $\beta$ at which $V_{\rm eff}(P_+) =V_{\rm eff}(P_-)$ are shown by blue curves in Fig.~\ref{fig5} (right).
The values of $\beta$ are indicated near the blue lines. 
Along the line for each $\beta$, the double-well potential appears, and the contour curve is expected to turn into the Stokes line in the infinite volume limit.
For finite volume, at points on that line, where the phase cancelation occurs, Lee-Yang zeros appear.
This result indicates the existence of singularities in the region of large $\mu_{\rm Im}$ 
as well as the region of large  $\mu_{\rm Re}$, and the boundary in the complex plane is closer
to the origin $\mu = 0$ than that (the square symbol) on the real axis.

\section{Conclusions}
\label{sec:summary}
We studied the distribution function and effective potential of QCD with two light flavors and $N_{\rm f}$ massive flavors, aiming to understand phase structure of 2 and (2+1)-flavor QCD.
Through the shape of the distribution function, we investigated the critical surface separating the first order transition and crossover regions in the parameter space of the light quark mass, heavy quark mass and the chemical potential.
It is found that the critical mass becomes larger with $N_{\rm f}$ and the first order region becomes wider with the increasing chemical potential for large $N_{\rm f}$. 
If (2+1)-flavor QCD has the same property, 
this gives the strong evidence for the existence of the critical point at
finite density in the real world.

The nature of the chiral phase transition in the 2-flavor massless limit is still open question.
To study the chiral limit, we investigated the light quark mass dependence. 
If the transition is of first order, the critical $\kappa_{\rm h}$ vanishes before going to the 2-flavor massless limit.
But, the critical $\kappa_{\rm h}$ does not show such a behavior in the region we investigated. 
This implies that the critical mass of heavy quark remains finite even in the chiral limit of 2-flavors and 2-flavor QCD near the chiral limit is not in the first order transition region.

We then discussed 2-flavor QCD with the complex $\mu$ by a numerical simulation. 
We found that there is a region where the plaquette distribution function has two peaks, suggesting the existence of singularities characterized by ${\cal Z}=0$, in the region of large $\mu_{\rm Im}$ as well as of large $\mu_{\rm Re}$. 
Combined with the analytical study of the complex chemical potential, the distribution of the singularities may provide important information about the QCD phase transition.

\end{document}